\documentclass[10pt,letterpaper]{article}
\usepackage[top=0.85in,left=2.75in,footskip=0.75in]{geometry}

\usepackage{amsmath,amssymb}

\usepackage{changepage}

\usepackage[utf8x]{inputenc}

\usepackage{textcomp,marvosym}

\usepackage[table]{xcolor}

\usepackage{cite}

\usepackage{todonotes}

\usepackage{nameref,hyperref}

\usepackage[right]{lineno}

\usepackage{microtype}
\DisableLigatures[f]{encoding = *, family = * }

\usepackage[table]{xcolor}

\usepackage{array}

\newcolumntype{+}{!{\vrule width 2pt}}

\newlength\savedwidth

\newcommand\thickhline{\noalign{\global\savedwidth\arrayrulewidth\global\arrayrulewidth 2pt}%
\hline
\noalign{\global\arrayrulewidth\savedwidth}}

\usepackage{setspace} 
\raggedright
\usepackage[aboveskip=1pt,labelfont=bf,labelsep=period,justification=raggedright,singlelinecheck=off]{caption}

\makeatletter
\renewcommand{\@biblabel}[1]{\quad#1.}
\makeatother

\usepackage{lastpage,fancyhdr,graphicx}
\usepackage{epstopdf}
\fancyheadoffset[L]{2.25in}
\fancyfootoffset[L]{2.25in}
\begin{document}
\vspace*{0.2in}

% Title must be 250 characters or less.
\begin{flushleft}
{\Large
\textbf\newline{Eye Movement Feature Classification \\for Soccer Goalkeeper Expertise Identification in Virtual Reality} % Please use "sentence case" for title and headings (capitalize only the first word in a title (or heading), the first word in a subtitle (or subheading), and any proper nouns).
}
\newline
% Insert author names, affiliations and corresponding author email (do not include titles, positions, or degrees).
\\
Benedikt W. Hosp \textsuperscript{*,1,2},
Florian Schultz\textsuperscript{2},
Oliver Höner\textsuperscript{2},
Enkelejda Kasneci\textsuperscript{1}

\bigskip
\textsuperscript{1} Human-Computer-Interaction, University of Tübingen, Germany
\\
\textsuperscript{2} Institute of Sport Science, University of Tübingen, Germany
\\
\bigskip

% Insert additional author notes using the symbols described below. Insert symbol callouts after author names as necessary.
% 
% Remove or comment out the author notes below if they aren't used.
%

% Additional Equal Contribution Note
% Also use this double-dagger symbol for special authorship notes, such as senior authorship.
%\ddag These authors also contributed equally to this work.

% Current address notes
%\textcurrency Current Address: Dept/Program/Center, Institution Name, City, State, Country % change symbol to "\textcurrency a" if more than one current address note
% \textcurrency b Insert second current address 
% \textcurrency c Insert third current address

% Deceased author note
%\dag Deceased

% Group/Consortium Author Note
%\textpilcrow Membership list can be found in the Acknowledgments section.

% Use the asterisk to denote corresponding authorship and provide email address in note below.
* benedikt.hosp@uni-tuebingen.de (BH)

\end{flushleft}
% Please keep the abstract below 300 words
\section*{Abstract}
The latest research in expertise assessment of soccer players has affirmed the importance of perceptual skills (especially for decision making) by focusing either on high experimental control or on a realistic presentation. To assess the perceptual skills of athletes in an optimized manner, we captured omnidirectional in-field scenes and showed these to 12 expert, 10 intermediate  and 13 novice soccer goalkeepers on virtual reality glasses. All scenes were shown from the same natural goalkeeper perspective and ended after the return pass to the goalkeeper. Based on their gaze behavior we classified their expertise with common machine learning techniques. This pilot study shows promising results for objective classification of goalkeepers expertise based on their gaze behaviour and provided valuable insight to inform the design of training systems to enhance perceptual skills of athletes.

%\linenumbers

\section{Introduction}
\label{sec:intro}

In addition to physical performance factors, perceptual-cognitive skills play an increasingly important role as cognitive performance factors in sport games, too. In perceptual research examining the underlying processes of these skills, subjects are typically placed in a situation where they have to react while their behaviour is being recorded and subsequently analyzed. Such behaviour can be assigned to a class, for example, to provide information about performance levels. Many studies in sport games in general and in soccer in particular ~\cite{berry2008contribution,catteeuw2009decision,abernethy2010revisiting,helsen1999multidimensional,abernethy1999can,farrow2002can,williams2002anticipation} have shown that athletes in a high-performance class have a more highly developed perception, leading – amongst other factors – to success in sports. However, this research is still confronted with challenges regarding experimental control and a representative presentation of the situation. Furthermore, the potential of more and more used technologies such as eye-tracking systems to assess the underlying perceptual-cognitive processes has not yet been realized, especially with regard to the analysis of complex eye-tracking data. In this work, we research how to handle and analyze such a large and complex data in an optimized way by applying common supervised machine learning techniques to the gaze behaviour of soccer goalkeepers during a decision-making task in build-up game situations presented as 360°-videos in a consumer-grade virtual reality headset.

%2. Expertise in decision making
Latest sports-scientific expertise research shows that experts - when it comes to decision-making- have more efficient gaze behavior, because they apply an advanced cue utilization to identify and interpret relevant cues ~\cite{mann2007perceptual}. This behavior enables 
experts to make more efficient decisions than non-experts, e.g. during game build-up by the goalkeeper. From both a scientific and practical sports perspective, of particular importance are factors that lead to successful perception, form expertise and how these can be measured. 
% 2. Diagnostic model
To measure perception based expertise, at first a diagnostic system is needed for recognition of expertise, which provides well-founded information about the individual attributes of perception. These attributes are usually considered in isolation. Thus, their influence on expertise can be specifically recognized.
% Trade-off problem
To allow the athletes to apply their natural gaze behaviour, the experimental environment is important, but one of the main problems in perceptual cognitive research persists: realism vs. control. In a meta review of more than 60 studies on natural gaze behavior from the last 40 years, Kredel et al. \cite{kredel2017eye} postulate that the main challenges in perception research lie in a trade-off between experimental control and a realistic valid presentation. Diagnostic and training models are often implemented or supported by digital means.
% These are considered popular and ideal for perception research. For example,  Larkin et al. \cite{larkin2015evaluation} concluded from 25 studies that video-based training methods can improve perceptual-cognitive performance. 
 
% VR lösung
This is nothing new, as  in sports psychological research, new inventions in computer science such as presentation devices (i.e. CAVE \cite{defanti2009starcave}, virtual reality (VR) \cite{wirth2018assessment}), interface devices (i.e. virtual reality, leap motion etc.) or biometric feature recognition devices (i.e. eye tracker \cite{nystrom2018tobii}) are used more and more often. 
% VR ist immersive und realistic
As a new upcoming technology, virtual reality (VR) devices are used more frequently as stimulus presentation and interaction devices. As said, a fundamental aspect in perception research is a highly realistic presentation mode, which allows for natural gaze behavior during diagnostic. VR technology makes this possible by displaying realistic, immersive environments. However, this strength, allowing natural gaze behaviour, comes less from the VR technology itself. According to Gray R. \cite{gray2019virtual} it depends on psychological fidelity, the degree to which the perceptual-cognitive requirements of the real task are replicated in such environments. Next to immersion and presence, Harris et al. \cite{harris2020framework} suggest the expansion of a simulation characterization into a typology of fidelity (containing also psychological fidelity) to determine the realism of a simulation.  VR offers an immersive experience through the use of 4k 360\textdegree video, which experiences a higher level of realism than, for example, CAVE systems, by providing higher levels of psychological fidelity ~\cite{gray2019virtual,harris2020framework}. VR is therefore a popular and optimal tool for perception research. Bideau et al. \cite{bideau2010using} summarize further advantages of VR in their work. Their main contribution, however, is their immersive virtual reality that elicits expert responses similar to real-world responses.

In a narrower sense, VR is based on computer generated imagery (CGI). One advantage of such fully CGI-based environments is the possibility of the user interacting with the environment, which presumingly increases the immersive experience. On the other hand, fully CGI-based environments contain moving avatars that are from natural in appearance and hides environmental influences. This might prevent high immersion and influence the participant's gaze behaviour. Therefore, we chose a realistic environment with 360\textdegree stimuli to provide a close to natural environment that does not influence the participant's gaze behaviour. As this work presents an focus on the cognitive processes of decision-making, we focus less on realistic interaction methods

Especially interesting are the developments of VR devices regarding integrated measuring devices. More and more devices have  eye trackers directly integrated, which, in combination with a photo-realistic environment in VR glasses, allows for the measurement of almost optimal user gaze behavior while also showing highly realistic stimuli.  Eye trackers provide a sound foundation with high temporal and spatial resolution to research perceptual processes. The combination of VR and high speed eye tracking allows the collection of a massive amount and highly complex data. With the high quality eye images and freedom of movement of an mobile eye tracker, the high speed of a remote eye tracker and the the control over the stimulus in lab setting (VR) and naturality of in-situ stimuli by omnidirectional videos, the outcome of this combination is highly complex. Analysis of such data is a particular challenge, which emphazises the need for new analysis methods.
As we want to infer underlying mechanisms of perceptual-cognitive expertise, tracking eye movements is our method of choice in this work. Generally, perceptual research focuses on eye tracking because, as a direct measuring method, it allows for a high degree of experimental control. Besides a realistic presentation and high degree of experimental control, VR can also be used to model the perception \cite{duchowski2000binocular} of athletes and thus creates a diagnostic system. A diagnostic system has the ability to infer the current performance status of athletes to identify performance-limiting deficits, an interesting provision of insight for the athletes and coach as well. Most importantly, such a diagnostic system forms the basis for an adaptive, personalized and perceptual-cognitive training system to work on the reduction of these deficits.

% Übergang zu Eye tracking
So far, eye tracking studies have focused on either in-situ setups with realistic presentation mode and mobile eye trackers (field camera showing the field of view of the user) or on laboratory setups with high experimental control using remote eye trackers ~\cite{uppara2018eye,grushko2014usage,bard1981considering,bahill1984can,singer1998new,vickers1992gaze}. Since mobile eye trackers are rarely faster than 100-120 Hz, because saccades and smooth pursuits cannot be detected properly at such speed, investigations in an in-situ context are limited to the observation of fixations. Fixations are eye movement events during which the eye is focused on an object for a certain period of time (and thus projects the object onto the fovea of the eye) so that information about the object can be cognitively processed. The calculation of fixations with such a slow input signal leads to inaccuracies in the recognition of the start and end of the fixation. 
Only limited knowledge can be gained using such eye trackers because additional information contained in other eye events, such as saccades and smooth pursuits, cannot be computed correctly. This prevents the use of such eye trackers as robust expert measurement devices. Saccades are the jumps between the fixations that allow the eye to realign. They can be as fast as 500 \textdegree/s. Smooth pursuits are especially interesting in ball sports because they are fixations on moving objects i.e. moving players. However, especially in perception studies in soccer in VR-like environments, slow eye trackers with about 25-50 Hz are primarly used \cite{roca2020perceptual,dicks2010examination,aksum2020football,roca2018creative}. This speed limits the significance of these studies to fixation and attention distribution in areas of interest (AOI). Aksum et al. \cite{aksum2020football}, for example, used the Tobii Pro Glasses 2 with a field camera set to 25 Hz. Therefore, only fixations or low speed information is available and no equal stimuli for comparable results between participants. In a review of 38 studies, McGuckian et al. \cite{mcguckian2018systematic} summarized the eye movement feature types used to quantify visual perception and exploration behaviour of soccer players. Except for Bishop et al.  \cite{bishop2014telling}, all studies were restricted to fixations thus restricting the gainable knowledge of eye movement features. The integration of  high speed eye trackers into VR glasses combines both strengths: high experimental control of a high speed eye tracker and a photo-realistic stereoscopic VR environment.

%3.	Machine learning als Analyse methode
With more frequent use of eye trackers, and more accurate, faster and ubiquitous devices, huge amounts of precise data from fixations, saccades and smooth pursuits can be generated which cannot be handled in entirety utilizing previous analysis strategies. Machine learning provides the power to deal with huge amounts of data. In fact, machine learning algorithms typically improve with more data and allow - by publishing the model's parameter set - fast, precise and objective reproducible ways to conduct data analysis. Machine learning methods have already been successfully applied in several eye tracking studies. Expertise classification problems in particular, can be solved as shown by Castner et al. in dentistry education ~\cite{castner2020deep,castner2018scanpath} and Eivazi et al. in microsurgery ~\cite{bednarik2013computational,eivazi2012gaze,eivazi2011predicting,eivazi2017towards}.  Machine learning techniques are the current state-of-the-art for expertise identification and classification. Both supervised learning algorithms ~\cite{bednarik2013computational,castner2018scanpath} and unsupervised methods or deep neural networks ~\cite{castner2020deep} have shown their power for this kind of problem solving. This combination of eye tracking and machine learning is especially well suited when it comes to subconscious behavior like eye movements features as these methods have the potential to greatly benefit the discovery of different latent features of gaze behavior and their importance and relation to expertise classification. 

%5.	Contribution
In this work, we present a model for objective recognition of soccer goalkeepers' expertise in regard to decision-making skills in build-up situations by means of machine learning algorithms relying solely on eye movements. We also present an investigation of the influences of single features on explainable differences between single classes. This pilot study is meant to be a first step towards a perceptual-cognitive diagnostic system and a perceptual-cognitive virtual reality training system, respectively.

\section{Method}

%The basis of this work is a pilot study on a VR system with an integrated eye tracker. This chapter describes the experimental setup, the pilot study, the eye tracking characteristics and the methodical procedure for the analysis with machine learning methods.

\subsection{Experimental setup}
%\label{sec:sysdesc}
%\input{Chapters/project.tex}

In the present study we used an HTC Vive, a consumer-grade virtual reality (VR) headset. Gaze can be recorded through integration of the SMI high speed eye tracker at 250 Hz. The SteamVR framework is an open-source software that interfaces common real-time game engines with the VR glasses to display custom virtual environments. We projected omnidirectional 4k footage on the inside of a sphere that envelopes the user's field of view, which leads to high immersion in a realistic scene.

\subsubsection{Stimulus material}

We captured the 360\textdegree -footage by placing an Insta Pro 360 (360\textdegree camera) on the soccer field on the position of the goalkeeper. Members of a German First League's elite youth academy were playing 26 different 6 (5 field players + goalkeeper) versus 5 match scenes on one half of a soccer field. Each scene was developed with a training staff team of the German Football Association (DFB) and each decision was ranked by this team. There were 5 options (teammates) plus one "emergency" option (kick out). For choosing the option rated as the best option by the staff team, the participant earned 1 point, because this option is the best option to ensure continuation of the game. All other options were rated with 0 points. Conceptually, all videos had the following content: The video starts with a pass by the goalkeeper to one of the teammates. The team passes the ball a few times until the goalkeeper (camera position) receives the last return pass. The video stops after this last pass and a black screen is presented. The participant now has 1.5 seconds time to report which option they've decided on and the color of the ball which was printed on the last return pass (to force all participants to recognize the last return pass realistically).

\subsubsection{Participants}

We collected data from 12 german expert youth soccer goalkeepers (U-15 to U-21) during two youth elite goalkeeper camps. The data from 10 intermediates was captured in our laboratory and comes from regional league soccer goalkeepers (semi-professional). Data from 13 novices came from players with up to 2 years of experience with no participation in competitions and no training on a weekly basis. The experts have 8.83 hours training each week and are 16.6 years old on average. They actively playing soccer for about 9 years, which is significantly more than the novices (1.78 years), but less than the intermediates (15.5 years). This may be a result of their age difference. The intermediates are 22 years old on average, but have nearly half of the training hours per week compared to the experts. Characteristics of the participants can be seen in Table ~\ref{tab:participants}.

\begin{table}[!h]
	\renewcommand*{\arraystretch}{1.6}
	\caption{Participants summary.}
	~\label{tab:participants}
	\centering
	\begin{tabular}{l l r r r}
		% \toprule
		& & \multicolumn{2}{c}{\small{\textbf{Participants}}} \\
		
		{\small\textit{Class}}
		& {\small \textit{Attribute}}
		& {\small \textit{Average}}
		& {\small \textit{Std. Dev.}}\\
		
		{Experts} & Age 				& 16.60 & 1.54  \\
		& Active years	    & 9.16 & 5.04   \\ 
		&Training hours/week  & 8.83 & 4.27  \\	
		\hline 
		
		{Intermediates} &Age			 	 & 22.00  & 3.72   \\
		&Active years	 	 & 15.50& 5.77  \\
		&Training hours/week & 4.94& 0.91  \\
		
		\hline 
		{Novices} &Age  				& 28.64 	& 3.72   \\
		&Active years 		& 1.78   	& 5.21 \\
		&Training hours/week  & 0.00 	    & 0.00  \\
		
		\hline 
	\end{tabular}
	
\end{table}

\subsubsection{Procedure}

The study was confirmed by the Faculty of Economics and Social Sciences Ethic Committee of the University of Tuebingen. After signing a consent form to allow the usage of their data, we familiarized the participants with the footage.

\begin{figure*}
	\centering
	\includegraphics[width=1\linewidth]{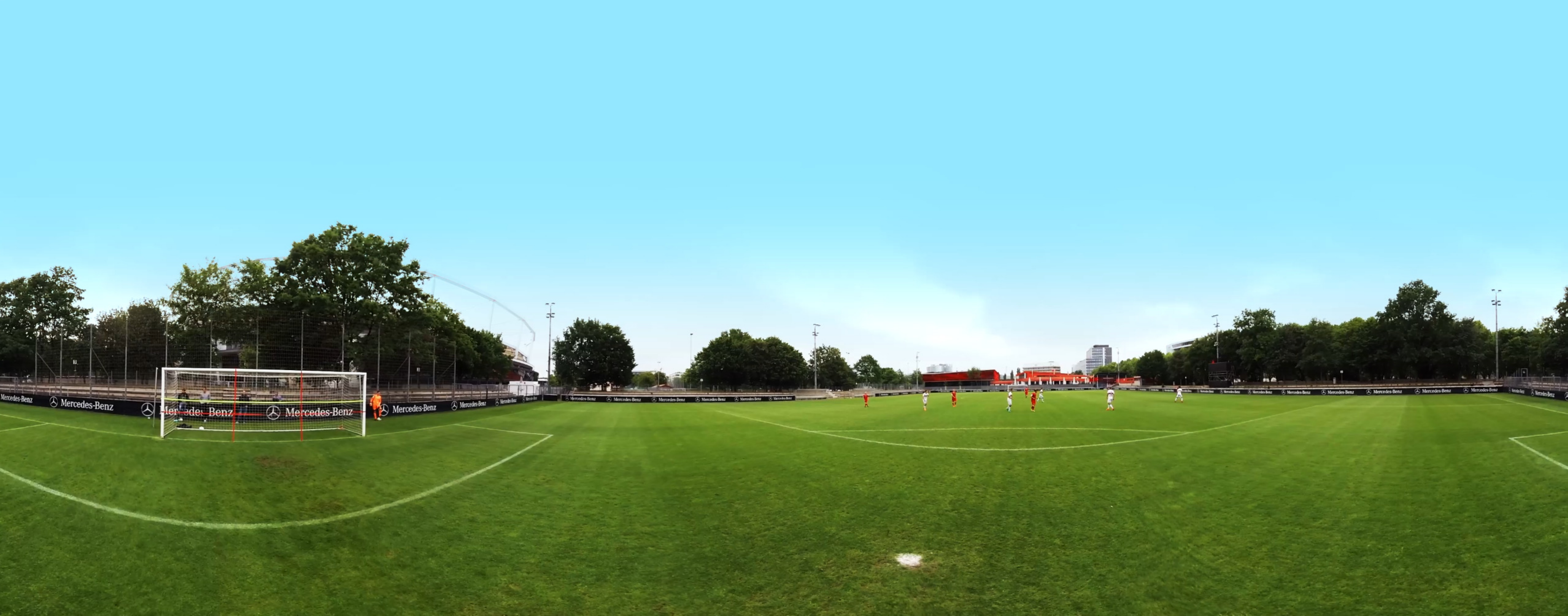}
	\caption{Example stimulus in equirectangular format. }~\label{fig:stimulus}
\end{figure*}

\begin{figure*}
	\centering
	\includegraphics[width=1\columnwidth]{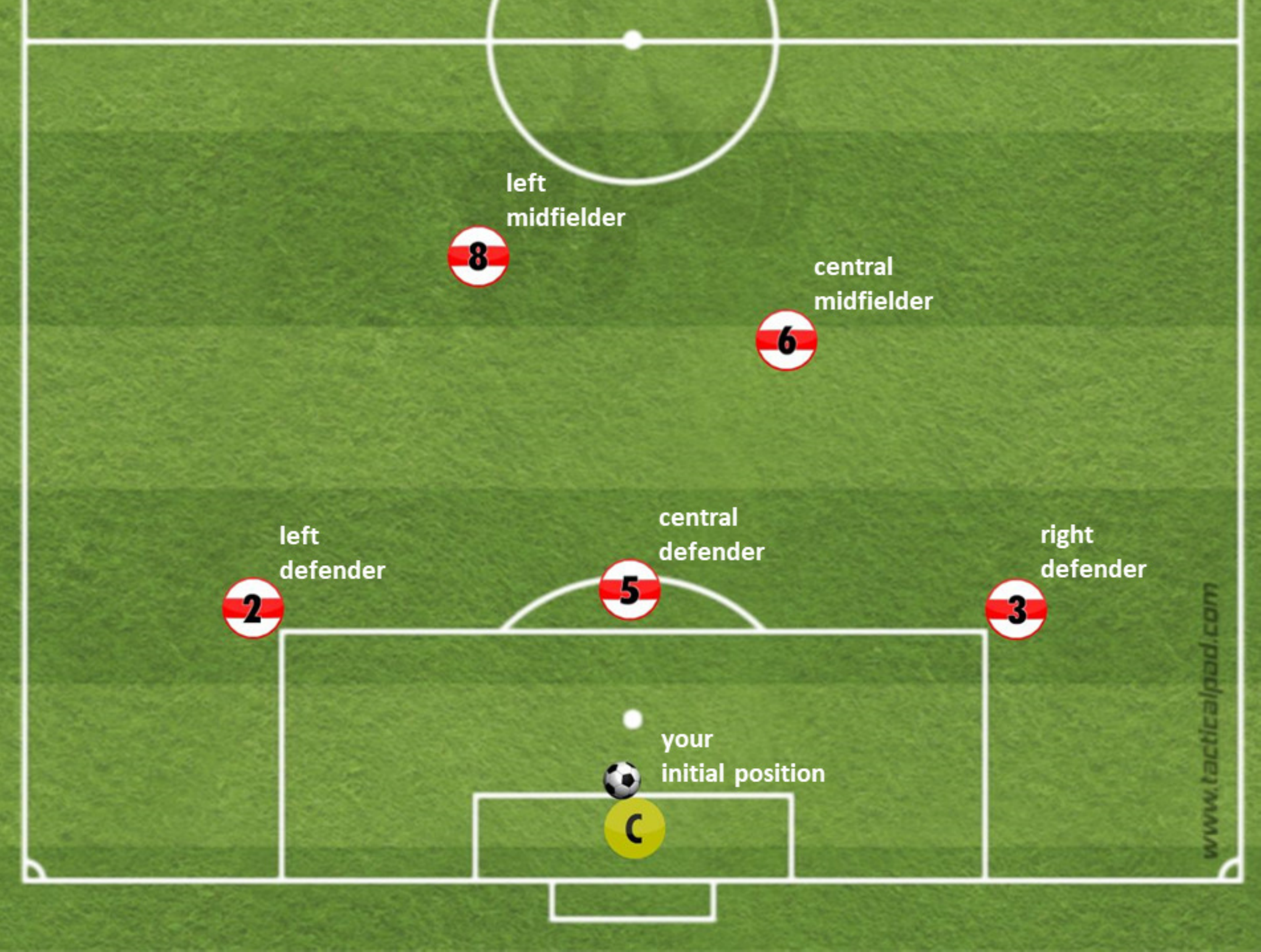}
	\caption{Schematic overview of the response options. Emergency option kick out is not shown. }
	~\label{fig:schematic}
\end{figure*}

The study contained two blocks consisting of the same 26 stimuli in each (conceptually as mentioned in the stimulus material section). The stimuli in the second block were presented in a different randomized order. Each decision made on the continuation of a video has a binary rating, as only the best decision was counted as 1 (correct) while all other options were rated as 0 (incorrect). At first, 5 different sample screenshots (example view see Fig ~\ref{fig:stimulus} in equirectangular form or S1 Video ~\ref{S1_Vid} for a cross section of the stimulus presentation sphere) and the corresponding sample stimuli were shown and explained to acclimate the participant to the setup. To learn the decision options, we also showed a schematic overview before every sample screenshot (see Fig ~\ref{fig:schematic}).

\subsection{Eye tracking}
\label{sec:method}
The raw data of the SMI Eye tracker can be exported from the proprietary BeGaze software as csv files. BeGaze already provides the calculation of different eye movement features based on the raw gaze points. As we get high speed data from the eye tracker, we use the built-in high speed event detection. The software first calculates the saccades based on the peak threshold, which means the minimum saccade duration (in ms) varies and is set dependent on the peak threshold default value of $40 ^{\circ}/s $. In a second step, the software calculates the fixations. Samples are considered to belong to a fixation when they are between a saccade or blink. With a minimum fixation duration of 50 ms we reject all fixations below this threshold. As there is no generally applicable method for detection of smooth pursuits, this kind of event is included and encoded as fixations with longer duration and wider dispersion. We marked fixations with a fixation dispersion of more than 100 px as smooth pursuits. By doing this, we split fixations into normal length fixations and long fixations which we consider to be and refer to smooth pursuits. This threshold is an empirical value based on the sizes of the players as main stimuli in the video. The following section describes the steps that are necessary to train a model based on these eye movement features.

%\todo{change Figure 2 to FOV or Sphere animation screenshot?}
\subsubsection{Feature selection}
\label{subfeaturesSelection}

As it is not clear which subset of eye movement features explains the difference in expertise completely, we follow a bruteforce method, considering all possible measures issued by the eye tracking device and subsequently evaluating their importance. For the classification of expertise level we focus on the following features:

\begin{itemize}
	\item  event durations and frequency (fixation/ saccade),
	\item fixation dispersion ( in \textdegree),
	\item smooth pursuit duration (in ms)
	\item smooth pursuit dispersion (in \textdegree)
	\item saccade amplitude (in \textdegree),
	\item average saccade acceleration (in \textdegree/$s^2$), 
	\item peak saccade acceleration (in \textdegree/$s^2$),
	\item average saccade deceleration (in \textdegree/$s^2$), 
	\item peak saccade deceleration (in \textdegree/$s^2$),
	\item average saccade velocity (in \textdegree/$s$),
	\item peak saccade velocity (in \textdegree/$s$).
	
\end{itemize}

Each participant viewed 26 stimuli twice, resulting in 52 trials per subject. First, we viewed the samples of these 52 trials and checked the confidence measures of the eye tracking device. We removed all trials with less than 75\% tracking ratio, as gaze data below this threshold is not reliable. Due to errors in the eye tracking device, not all participant data is available for every trial. Table \ref{tbl:erroneousTrials} shows an overview of the lost trials. For two participants, 11 trials had a lower tracking ratio; on participant 18, we lost 35 trials; and on participant 33, one trial was lost. This results in 1658 out of 1716 valid trials in total. 3.3\% of the trials were lost due to eye tracking device errors.
 
 \begin{table}[!h]
 	
% 	\begin{adjustwidth}{-2.25in}{0in} % Comment out/remove adjustwidth environment if table fits in text column.
 		\centering

 		\caption{{\bf Overview of the amount of erroneous trials, based on eye tracking device errors.}}
 		\begin{tabular}{|c|c|}
 			 		\hline
 			\multicolumn{2}{|c|}{\bf Overview erroneous trials} \\ 
 			Participant & Number of valid trials\\
 			\thickhline
 			 1 & 11\\	\hline		 		
 			 8 & 11\\\hline
			 18 & 25\\\hline
     		 33 & 1 \\\hline
     		all others & 0\\
		\hline
 		\end{tabular}
 		\label{tbl:erroneousTrials}
 		\begin{flushleft}\end{flushleft}
 		
 %	\end{adjustwidth}		
 \end{table}

\subsubsection{Data cleaning}
We checked the remaining data for the quality of saccades. This data preparation is necessary to remove erroneous and low quality data that comes from poor detection on behalf of the eye tracking device and does not reflect the correct gaze. Therefore, we investigated invalid samples and removed (1) all saccades with invalid starting position values, (2) all saccades with invalid intra-saccade samples, and (3) all saccades with invalid velocity, acceleration or deceleration values.

%72 of 17621
(1) Invalid starting position:  0.22\% saccades started at coordinates (0;0). This is an encoding for an error of the eye tracking device. As amplitude, acceleration, deceleration and velocity are calculated based on the distance from start- to endpoint, these calculations result in physiological impossible values, e.g., over 360\textdegree saccade amplitudes.

\begin{figure}
	\centering
	\includegraphics[width=1.0\columnwidth]{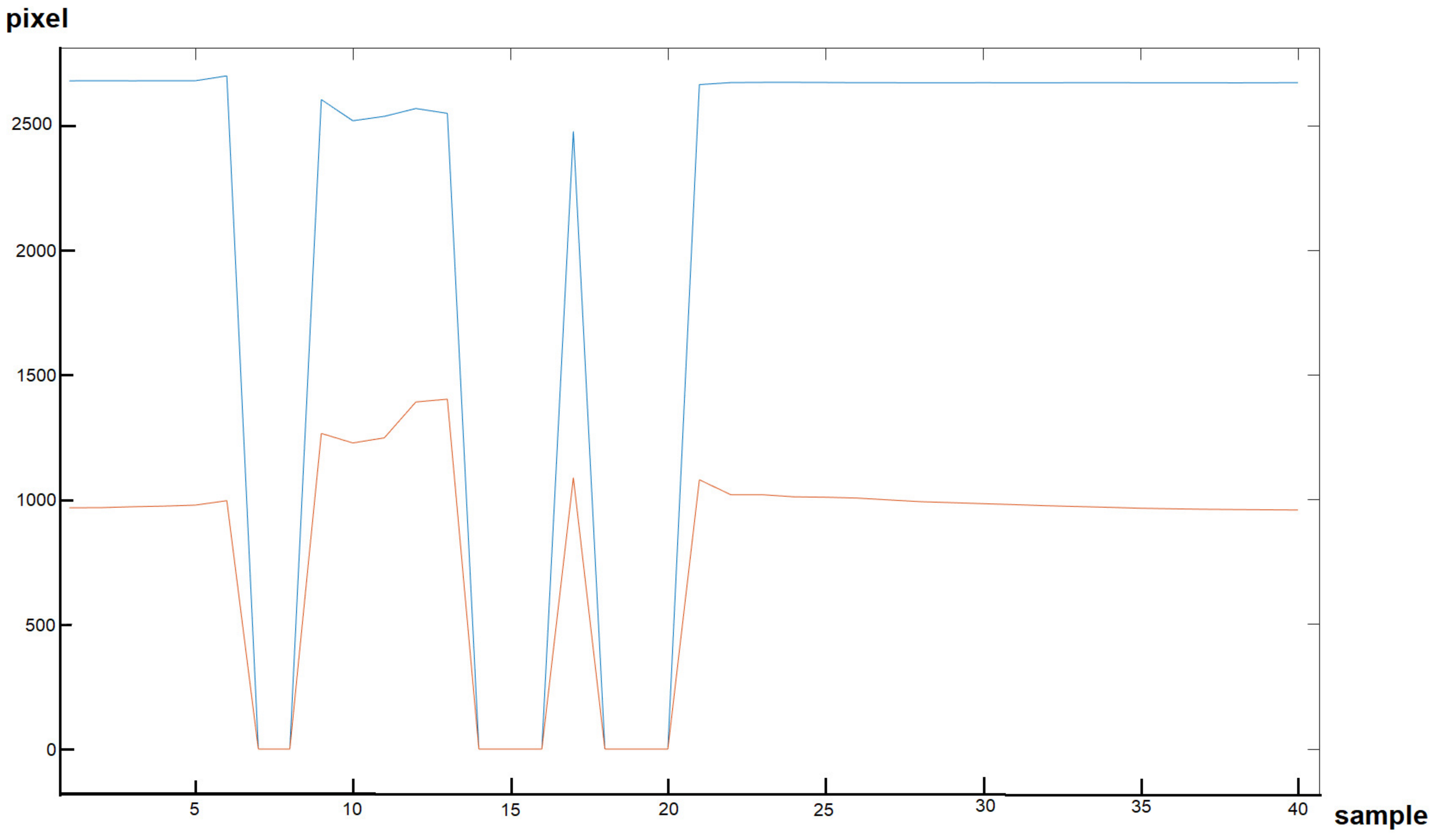}
	\caption{Example of invalid intra-saccade values. The x-axis shows the number of the gaze signal sample (40 samples, 250 Hz, 160 ms duration) and the y-axis shows the position in pixel. The blue line represents the x-signal of the gaze and the orange line the y-signal. }~\label{fig:intra-saccadeError}
\end{figure}

%359
(2) Invalid intra-saccade values: Another error of the eye tracking device stems from the way the saccade amplitude is calculated through the average velocity (Eq~\ref{eq:saccadeAmplitude}) which is based on the distance of the mean of start and endpoints on a sample-to-sample basis (see Eq ~\ref{eq:saccadeVelocity}). 3.6\% of the saccades had at least one invalid gaze sample and were removed (example see Fig \ref{fig:intra-saccadeError}).

\begin{eqnarray}
\label{eq:saccadeAmplitude}
\oslash Velocity * EventDuration
\end{eqnarray}

\begin{eqnarray}
 \label{eq:saccadeVelocity}
\frac{1}{n}*\sum_{1}^{n} \frac{dist(startpoint(i), endpoint(i))}{EventDuration(i)}
\end{eqnarray}

%\todo{bigger tick marks, axis labels, bigger values for fig. 3}

On Fig ~\ref{fig:intra-saccadeError}, the gaze signal samples 7, 8, 14-16, 18-20 (x-axis) both, the x- and y-signal (blue and red line, respectively) show zero values and thereby indicate a tracking loss. As the saccade amplitude is based on the average velocity which is calculated on a sample-to-sample Eq~\ref{eq:saccadeVelocity}, the velocity from samples 6 to 7, 8 to 9, 13 to 14, 16 to 17, 17 to 18, and 20 to 21 significantly increase the average velocity as the distances are high (on average over 2400 px for x-signal and over 1000px for y-signal, which corresponds to a turn of 225° on x-axis and 187.5° on y-axis in the time of 4 ms between two consecutive samples).

There are two interpretations for saccadic amplitude. The first refers to the shortest distance from start to end point of a saccadic movement (i.e., a straight line) and the second describes the total distance traveled along the (potentially curved \cite{holmqvist2011eye}, p.311) trajectory of the saccade. The SMI implementation follows the second definition. We could have potentially interpolated invalid intra-saccade samples instead of completely removing the complete saccade from analysis, however, this leads to uncertainties that can affect the amplitude depending on the amount of invalid samples and does not necessarily represent the true curvature of the saccade. 

%(3) Extreme amplitudes: As the vendor has not published how saccades are calculated when head movements are present, we restricted the data to saccades of an amplitude less than 60\textdegree. This value is based on the assumptions of  Guitton et al. \cite{guitton1987gaze} that a saccade is limited neurally to $\pm 53$\textdegree. In total 2.71\% of all saccades exceeded this threshold, possibly due to joint eye- and head movement.

(3) As the velocity increases as a function of the saccade amplitude \cite{collewijn1988binocular}, 4.8\% of the saccades were ignored because of the restriction on velocities greater than 1000\textdegree/s. Similar to extreme velocities, we removed all saccade samples that exceeded the maximum theoretical acceleration and deceleration thresholds. Saccades with longer amplitudes have higher velocity, acceleration and deceleration, but can not exceed the physiological boundaries of 100.000 \textdegree/$s^2$ \cite{holmqvist2011eye}. 3.0\% and 4.0\%, respectively, of all saccades that exceeded this limit. As most of the invalid samples had more than one error source, we only removed 5.5 \%  of the saccades (3.5\% of all samples) in total. 
%5.27% = FailedSamplesError / TotalSaccades

After cleaning the data we use the remaining samples to calculate the average, maximum, minimum and standard deviation of the features. This results in 36 individual features. We use those for classifying expertise in the following.

\subsection{Machine learning}
In the following, we refer to \textit{expert samples} as trials completed by an elite youth player of a DFB goalkeeper camp, \textit{intermediate samples} as those of regional league players and \textit{novice samples} as those of amateur players. We built a support vector machine model (SVM) and validated our model in two steps: cross-validation and leave-out validation. We trained and evaluated our model in 1000 runs with both validations. For each run, we trained a model (and validated with cross-validation) with samples of 8 experts, 8 intermediates, and 8 novices samples, and used the samples of two participants from each group of the remaining participants to predict their classes (leave-out validation). The experts as well as the intermediates and the novice samples in the validation set were picked randomly for each run.\\

\subsubsection{Sample assignment}

\begin{figure}[!h]
	\centering
	\includegraphics[width=1.0\columnwidth]{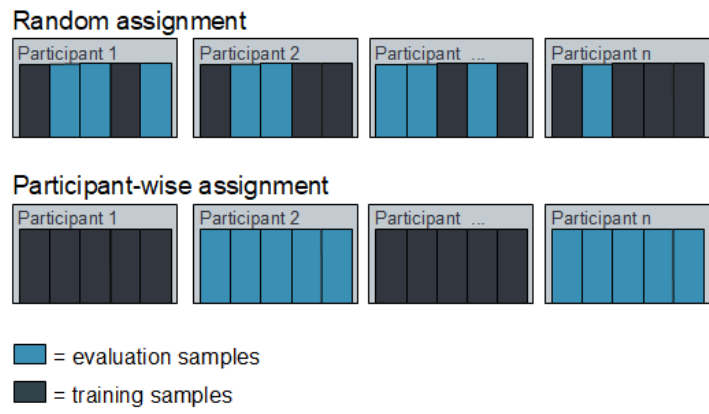}
	\caption{Example sample assignment. Top row shows a random assignment of samples, independent of the corresponding participant. Bottom row shows participant-wise sample assignment to training and evaluation set. }~\label{fig:sampleAssignment}
\end{figure}

We found that the way in which the data set samples are split into training and evaluation sets is very important and a participant-wise manner should be applied. By randomly picking samples independent of the corresponding participant, participant samples usually end up being distributed on the training and the evaluation set (illustrated in Fig  ~\ref{fig:sampleAssignment}). This leads to an unexpected learning behavior that does not necessarily classify expertise directly, but, rather, matches the origin of a sample to a specific participant thereby indirectly identifying that participants level of expertise. This means that a model would work perfectly for known participants, but is unlikely to work for unseen data. Multiple studies show that human gaze behavior follows idiosyncratic patterns. Holmqvist et al. \cite{holmqvist2011eye} show that a significant number of eye tracking measures underlay the participants' idiosyncrasy, which also means that the inter-participant differences are much higher than intra-participant differences. A classifier learns a biometric, person-specific measure instead of an expertise representation.

\subsubsection{Model building}
\begin{figure}
	\centering
	\includegraphics[width=1.0\columnwidth]{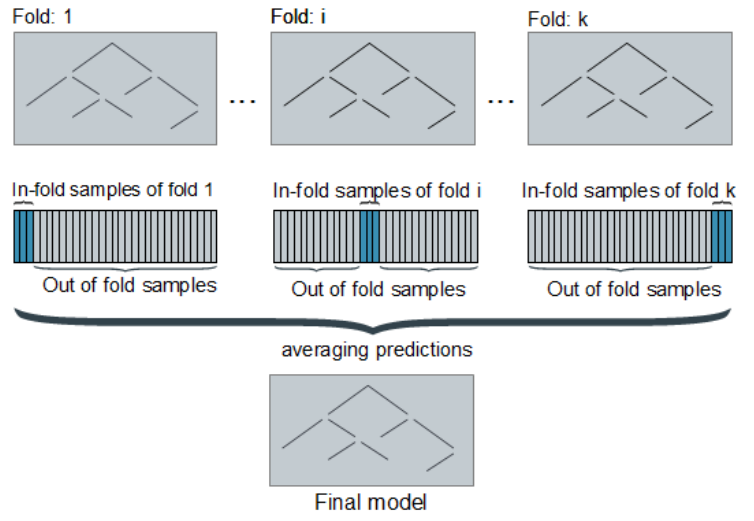}
	\caption{Illustration of the k cross-validation procedure. Each of the k models has a different out-of-fold and in-fold data set. We build the final model on the average of all predictions from all k models. }~\label{fig:trees}
	
\end{figure}

To find a model robust to high data variations, we applied a cross-validation during training. The final model is based on the average of k=50 models, with k = number of folds in the cross-validation. For each model $m_i$, with $i \in \{1, \dots ,k\}$, we use all out-of fold data of the i-th fold to train and evaluate $m_i$ with the in-fold data of the i-th fold (this procedure is illustrated in Fig ~\ref{fig:trees}). The final model is evaluated with a leave-out validation. The cross-validation step during training is independent from the leave-out validation with totally new data (never seen by the model). Information from cross-validation is used during the building and optimizing of the model and leave-out validation solely provides information about the prediction accuracy of the model when using completely new data. 

%\subsubsection{Prediction}

With a total of 810 valid samples, equally distributed on expert, intermediate and novice samples, we built a subset of 552 samples for training the model and a subset of 258 samples for evaluation. As each sample represents one trial, our approach here is to predict wether a trial belongs to an expert, intermediate or novice class. We tested assumption in different approaches.\\

\subsubsection{Classifiability}

Firstly, we used all 46 features to check the classifiability of this kind of data. The first approach contains all features from section \textit{Feature selection} \ref{subfeaturesSelection} with their derivations, (namely average, maximum, minimum, and standard deviation) to build an SVM  model (Tables \ref{tbl:novicesAll},\ref{tbl:intermediatesAll} and \ref{tbl:expertsAll} show all features with their derivations, splitted by class). When the binary case (expert vs. intermediates) results point out classifiability, the ternary case (expert vs. intermediate vs. novice) should be investigated.

\begin{table}[!h]
%	\begin{adjustwidth}{-2.25in}{0in} % Comment out/remove adjustwidth environment if table fits in text column.
		\centering
		\caption{
			{\bf All 46 features with their derivations. Novice class.}}
		\begin{tabular}{|l|c|c|c|c|}
			\hline
			\multicolumn{5}{|c|}{\bf Novices} \\  \hline
			
			Features & average & std. dev. & minimum & maximum  \\
			
			\thickhline
			\multicolumn{5}{|l|}{\bf Fixation} \\  \hline
		
			frequency (Hz)      & \cellcolor{gray!40}0.214  & - & - & -   \\ \hline
			duration (ms)       & \cellcolor{gray!40}214.017 & \cellcolor{gray!40}31.926 & \cellcolor{gray!40}190.49   &\cellcolor{gray!40} 239.30   \\ \hline
			dispersion (pixels) & \cellcolor{gray!40}72.092 &\cellcolor{gray!40}	25.68	& \cellcolor{gray!40}24.67  & 110.523 \\ \hline	
			
			\multicolumn{5}{|l|}{\bf Saccade} \\  \hline				
			frequency (Hz) & 0.071 & - & - & - \\ \hline
			duration (ms) & 71.688    &  \cellcolor{orange!70}38.869	& \cellcolor{orange!70} 26.514 & 175.460\\ \hline
			amplitude (\textdegree)&9.294 & 9.417 & \cellcolor{gray!40}0.574 &51.402 \\ \hline

			\multicolumn{5}{|l|}{\bf Saccade mean acceleration} \\  \hline				
			mean (\textdegree$/s^2$) &\cellcolor{gray!40} 4263.381 & 2482.019 & 366.666 & \cellcolor{gray!40}13984.563 \\ \hline
			peak (\textdegree$/s^2$) & \cellcolor{gray!40}9322.483168 & \cellcolor{gray!40}5777.273817 &\cellcolor{gray!40}231.836 & \cellcolor{gray!40}28355.224 \\ \hline
			
			\multicolumn{5}{|l|}{\bf Saccade deceleration } \\  \hline								
			peak (\textdegree$/s^2$)&\cellcolor{gray!40}-6848.104 &\cellcolor{orange!70} 4166.262 & \cellcolor{gray!40}-35563.646 & -411.760\\ \hline
			
			\multicolumn{5}{|l|}{\bf Saccade velocity } \\  \hline								
			mean (\textdegree$/s$) & \cellcolor{gray!40}105.463 & 65.023 &20.288 &298.134\\ \hline
			peak (\textdegree$/s$) & \cellcolor{gray!40}215.245 &  \cellcolor{orange!70}129.294 & 40.310 & \cellcolor{gray!40}766.157\\ \hline
			
			\multicolumn{5}{|l|}{\bf Smooth pursuit} \\  \hline								
			duration (ms)&\cellcolor{gray!40}302.637 & \cellcolor{gray!40}278.112& \cellcolor{gray!40}75.629 &\cellcolor{gray!40}1026.329\\ \hline	
			dispersion (pixels)& \cellcolor{orange!70}622.805 &201.268 & 	\cellcolor{orange!70}185.437 & 	\cellcolor{orange!70}1085.903\\ \hline

		\end{tabular}
		\label{tbl:novicesAll}
		\begin{flushleft} Gray cells show features with no significant differences between classes.
			Orange cells stand for a most frequent feature.
		\end{flushleft}
		
	%\end{adjustwidth}
	
\end{table}

\begin{table}[!ht]
	
%	\begin{adjustwidth}{-2.25in}{0in} % Comment out/remove adjustwidth environment if table fits in text column.
		\centering
		\caption{
			{\bf All 46 features with their derivations. Intermediate class.}}
		\begin{tabular}{|l|c|c|c|c|}
			\hline
			\multicolumn{5}{|c|}{\bf Intermediates} \\  \hline

			Features & average & std. dev. & minimum & maximum  \\
			\thickhline
			\multicolumn{5}{|l|}{\bf Fixation} \\  \hline
			frequency (Hz)      & \cellcolor{gray!40}0.255&	-	&-	&- \\ \hline
			duration (ms)       & \cellcolor{gray!40}255.225 &\cellcolor{gray!40}  53.379& \cellcolor{gray!40}215.835& \cellcolor{gray!40}299.623 \\ \hline
			dispersion (pixels) & \cellcolor{gray!40} 73.173 & \cellcolor{gray!40} 26.548  & \cellcolor{gray!40}23.070 & 114.762 \\ \hline		
			\multicolumn{5}{|l|}{\bf Saccade} \\  \hline				
			frequency (Hz) & 0.084&	-&	-& -\\ \hline
			duration (ms) & 84.349 & \cellcolor{orange!70} 59.726 & \cellcolor{orange!70}26.127  & 246.121\\ \hline
			amplitude (\textdegree)&9.883 & 10.674 &\cellcolor{gray!40} 0.572 &54.835\\ \hline
			\multicolumn{5}{|l|}{\bf Saccade mean acceleration} \\  \hline				
			mean (\textdegree$/s^2$) & \cellcolor{gray!40}4123.970 & 2685.991 & 315.346 & \cellcolor{gray!40} 15472.889 \\ \hline
			peak (\textdegree$/s^2$) & \cellcolor{gray!40}8920.177 & \cellcolor{gray!40} 5989.251 & \cellcolor{gray!40}216.722 & \cellcolor{gray!40}28266.000 \\ \hline
			\multicolumn{5}{|l|}{\bf Saccade deceleration } \\  \hline								
			peak (\textdegree$/s^2$)&\cellcolor{gray!40}-6948.491 & \cellcolor{orange!70} 4770.063 & \cellcolor{gray!40}-36334.137& -231.355 \\ \hline
			\multicolumn{5}{|l|}{\bf Saccade velocity } \\  \hline								
			mean (\textdegree$/s$) &\cellcolor{gray!40} 104.199 & 	66.682 & 21.520 & 331.111 \\ \hline
			peak (\textdegree$/s$) & \cellcolor{gray!40}213.835 &\cellcolor{orange!70} 136.529 & 40.109 & \cellcolor{gray!40}764.027 \\ \hline
			\multicolumn{5}{|l|}{\bf Smooth pursuit} \\  \hline								
			duration (ms)& \cellcolor{gray!40}291.092 & \cellcolor{gray!40}278.718 & \cellcolor{gray!40}73.835 & \cellcolor{gray!40}977.120 \\ \hline	
			dispersion (pixels)&  \cellcolor{orange!70}425.089 &  124.853 &  \cellcolor{orange!70}168.320 &  \cellcolor{orange!70}694.370 \\ \hline	

		\end{tabular}
		\label{tbl:intermediatesAll}
		\begin{flushleft} We consider samples as belonging to a smooth pursuit when the dispersion of the samples is greater than 100 px. As the size of the players in the stimulus varies around 90 pixel + a buffer.\end{flushleft}
%	\end{adjustwidth}
\end{table}

\begin{table}[!ht]
	
	%\begin{adjustwidth}{-2.25in}{0in} % Comment out/remove adjustwidth environment if table fits in text column.
		\centering
		\caption{
			{\bf All 46 features with their derivations. Expert class.}}
		\begin{tabular}{|l|c|c|c|c|}
			
			\hline
			\multicolumn{5}{|c|}{\bf Experts} \\  \hline

			Features & average & std. dev. & minimum & maximum  \\

			\thickhline
			\multicolumn{5}{|l|}{\bf Fixation} \\  \hline
			frequency (Hz)&\cellcolor{gray!40} 0.241 &	-& 	- & - \\ \hline
			duration (ms)& \cellcolor{gray!40} 241.509 & \cellcolor{gray!40} 58.629 & \cellcolor{gray!40}198.132 & \cellcolor{gray!40} 291.721 \\ \hline
			dispersion (pixels)&  \cellcolor{gray!40} 72.837 &\cellcolor{gray!40}	25.989&\cellcolor{gray!40}21.736 & 114.549\\ \hline		
			\multicolumn{3}{|l|}{\bf Saccade} \\  \hline				
			frequency (Hz)& 0.007 & - &- &- \\ \hline
			duration (ms) &65.472 &  \cellcolor{orange!70}35.548 &\cellcolor{orange!70} 25.019 & 163.415 \\ \hline
			amplitude (\textdegree)&  8.938 & 9.430 & \cellcolor{gray!40} 0.567 & 52.029\\ \hline
			\multicolumn{3}{|l|}{\bf Saccade mean acceleration} \\  \hline				
			mean (\textdegree$/s^2$) &\cellcolor{gray!40} 4769.655 & 3064.343 & 390.094 & \cellcolor{gray!40} 18965.944 \\ \hline
			peak (\textdegree$/s^2$) &\cellcolor{gray!40} 10026.456 & \cellcolor{gray!40} 7094.930 & \cellcolor{gray!40}175.242 & \cellcolor{gray!40}39445.125\\ \hline
			\multicolumn{3}{|l|}{\bf Saccade deceleration } \\  \hline								
			peak (\textdegree$/s^2$)& \cellcolor{gray!40}-7912.190 & \cellcolor{orange!70} 5492.287 &\cellcolor{gray!40} -43479.916 & -362.396\\ \hline
			\multicolumn{3}{|l|}{\bf Saccade velocity } \\  \hline								
			mean (\textdegree$/s$) & \cellcolor{gray!40} 110.675 & 72.737 & 21.182 & 375.363 \\ \hline
			peak (\textdegree$/s$) &  \cellcolor{gray!40}238.371 & \cellcolor{orange!70} 157.740 & 40.262 &\cellcolor{gray!40} 935.514 \\ \hline
			\multicolumn{3}{|l|}{\bf Smooth pursuit} \\  \hline								
			duration (ms) & \cellcolor{gray!40}276.785 &\cellcolor{gray!40} 265.679 & \cellcolor{gray!40}74.404 & \cellcolor{gray!40}953.660 \\ \hline	
			dispersion (pixels)&   \cellcolor{orange!70}399.939 & 112.414 &  \cellcolor{orange!70}336.016 & \cellcolor{orange!70} 505.031 \\ \hline	

		\end{tabular}
		\label{tbl:expertsAll}
		\begin{flushleft}\end{flushleft}
		
	%\end{adjustwidth}		
\end{table}

\subsubsection{Significant features}
\label{subsubsec:wilcoxonFeatures}
Secondly, we had a look at the features themselves and check for differences between the single features according to their class and as well as checking for the significance level of feature differences under 0.11\%. We built a model based on the features that have a significance level under 0.11\% (Tables ~\ref{tbl:novicesAll}, \ref{tbl:intermediatesAll} and \ref{tbl:expertsAll} all white cells, gray cells mean there is no significant difference between the groups).

\subsubsection{Most frequent features}
\label{subsubsec:mostfrequentfeatures}
In a third approach, we reduced the number of features by running the prediction on all 46 features 1000 times. By taking the most frequent features in the model, we search for a subset of features that prevent the model from overfitting and allow for interpretable results representing the differences between expertise classes with a minimum amount of features. These most frequenct features are imperative for the model to distinguish the classes. During training, the model indicates which features are the most important for prediction in each run. The resulting features with the highest frequency (and therefore highest importance for the model) in our test can be seen in Tables ~\ref{tbl:novicesAll},~\ref{tbl:intermediatesAll} and ~\ref{tbl:expertsAll}, in orange.

\section{Results}
\label{sec:results}

We first report the results of an intra-expert classification test to see whether inter-experts differences are smaller than than inter-class differences. Then, since we first need to know whether there are differences between experts and novices, the classifiablity test (binary classification) provides a deeper analysis on the model trained with all features for distinguishing experts and novices. The remaining chapter describes two ternary models which are based on a subset of features obtained through 1) their significance level and 2) their frequency in the all feature model. 

\subsubsection{Intra-expert classification}
To strengthen the implicit assumption of this paper that it is possible to distinguish between novices, intermediates and experts based on their gaze behavior, we evaluated our expert data separately by flipping a subset of experts with intermediates. After 100 iterations in which half of the experts where randomly labeled as intermediates, the average classification accuracy was below chance-level, meaning the model can not differentiate between experts and flipped experts properly. This strengthens our assumption that inter-expert differences are smaller than inter-group differences between experts, intermediates and novices.

\subsubsection{Binary classification}
\begin{figure}
	\centering
	\includegraphics[width=0.5\columnwidth]{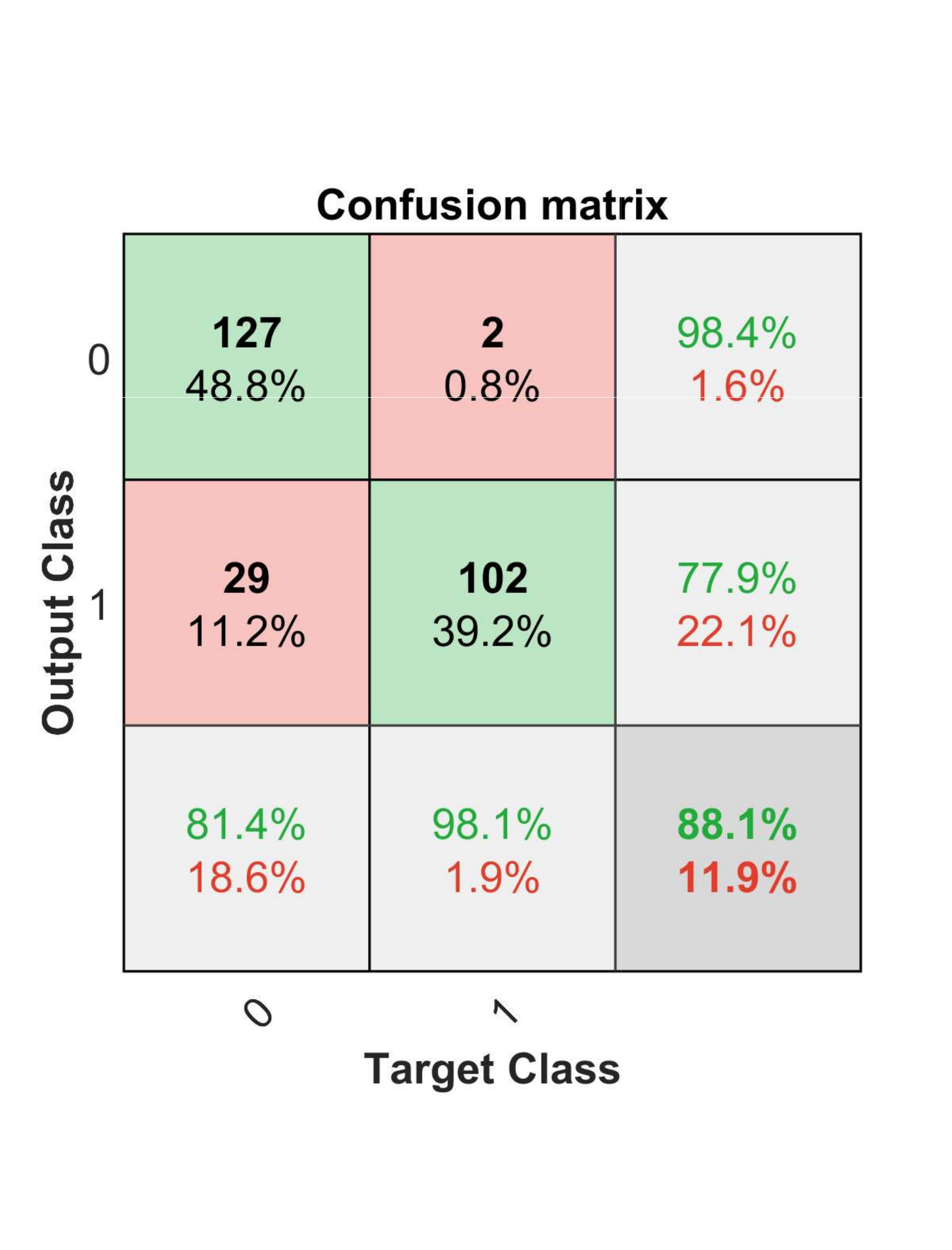}
	\caption{Binary confusion matrix about predictions on 100 randomized runs. }~\label{fig:confusionMFF}	
\end{figure}

The classifiability test shows promising results. This binary model is able to distinguish between experts and intermediates with an accuracy of 88.1\%. The model has a false negative rate of 1.6\% and a false positive rate of 18.6\%. This means the binary model predicted two out of 260 samples falsely as class zero and 29 samples that are class zero as class one. As the false negative rate is pretty low, the resulting miss rate is low (11.9\%) as well. The confusion matrix (Fig ~\ref{fig:confusionMFF}) shows the overall metrics. The binary model is better in predicting class zero samples (intermediates) than class one samples (experts). The overall accuracy of 88.1\% is sufficient to investigate on a ternary classification. In the following, we show deeper insights on the ternary approaches by looking at accuracy, miss rate and recall of the ternary models and compare those values between the All-feature model (ALL), most frequent features model (MFF) and significant features model (SF). This is to see if there is a better performing model with less features.

%The following features had the highest importance values on 100 runs: average fixation count, average fixation dispersion, standard deviation of fixation dispersion and standard deviation of saccade duration. As some features are ranked more than once, we only investigate on 4 features with highest importance value. 
%This reduction of features leads (1) to a slightly better overall performance on unseen data (median: 83\% to 88.4\%, see figure \ref{fig:boxplot_mff}) for the binary case but to lower overall performance for ternary case (median: 88\% to 73\%), and (2) to less deviation of accuracy in binary case and more in ternary case (lower adjacent, binary: 46\% to 65\%; lower adjacent, ternary: 39\% to 33\%). For the binary case, the whole performance increased, as the lower adjacent is now higher than 65\%. The IQR ranges from 80.69\% to 97.4\%. The median accuracy rises to 88.4\%. This means that the reduction of features improves the overall performance as well as it allows the interpretation of the used features in binary case. Figure \ref{fig:confusionMFF} shows an example confusion matrix for the binary case. The reduction to only most important features in the ternary case leads to an accuracy drop by 9\%. 

\subsection{Accuracy}
\label{subsubsec:classifiability}

\begin{figure}
	\centering
	\includegraphics[width=1\columnwidth]{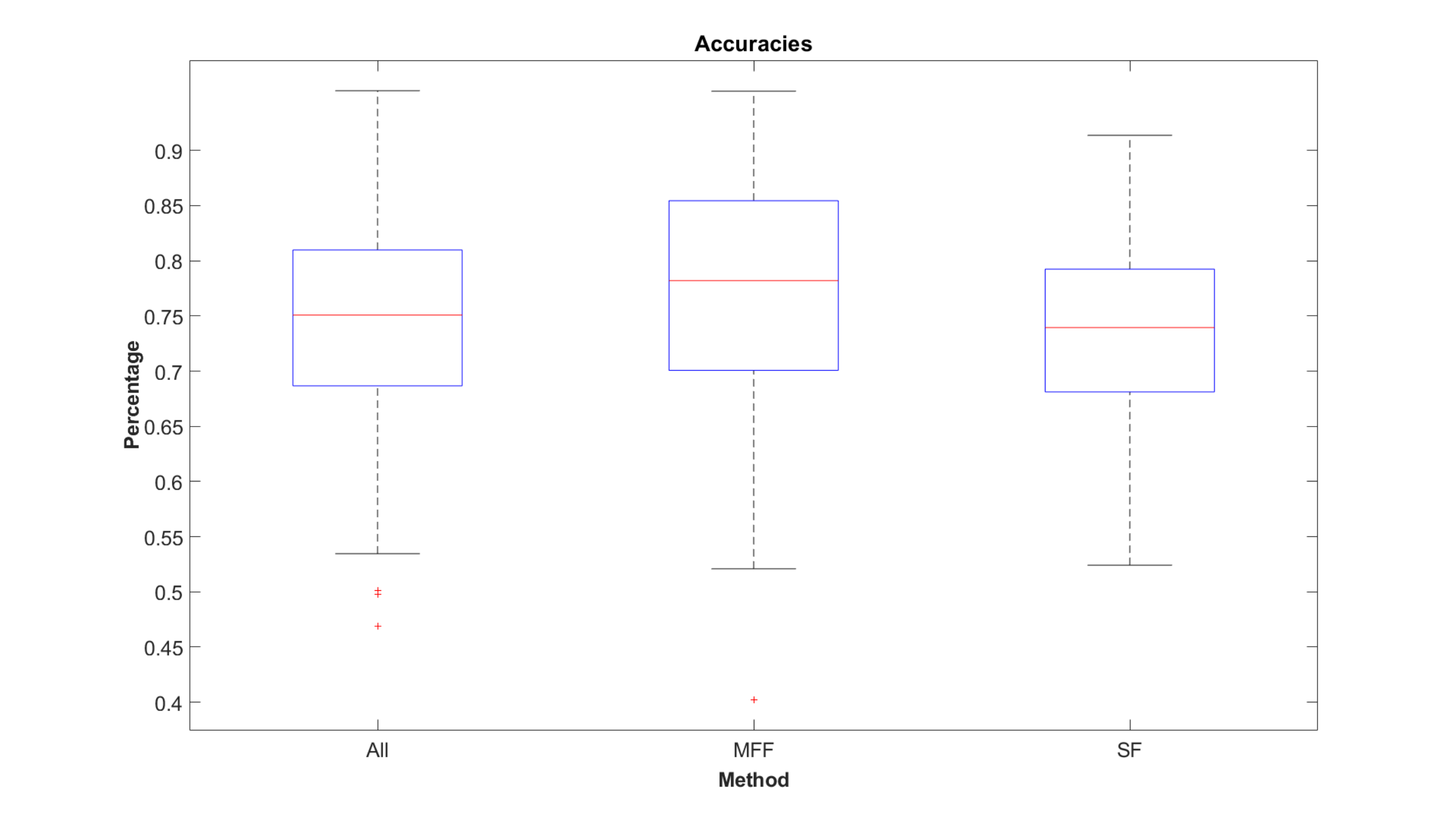}
	\caption{Boxplot showing the accuracy values of the ternary methods. All three models have median accuracy values $\sim 75-80\%$}
	~\label{fig:accuracies}
	
\end{figure}

The differences in accuracy between the three approaches are barely visible when looking at the median (ALL: 75.08\%, MFF: 78.20\%, SF: 73.95\%), but even greater when comparing the 75th percentile (ALL: 80.989\%, MFF: 85.44\%, SF: 79.25\%, see Fig\ref{fig:accuracies}). All models show a wider range of accuracy values which means these models might overfit more on some runs and underfit on others. The lower adjacent of all models is higher than chance level (ALL: 53.46\%, MFF: 52.93\% and SF:52.41\%), which means all models perform better as guessing. The chance level for 3 classes is 33.33\%. A system that would only guess the correct class would usually end up with an accuracy of about 33.33\%. Although not in each run, on average all models show a much better performance. Even the worst classification is over 20\% higher than chance level. A successful performance for classification expertise in machine learning models is usually when their average accuracy is between 70\% and 80\%. A statement about the performance of a model with lower than 70\% accuracy depends on the task and how much data is available. Sometimes there are only a few people in the world who can be considered experts. As the accuracy is a rough performance metric which only provides information about the number of correct predictions (true positives and true negatives), we offer a more detailed look into the performance of the methods by comparing the miss rates of the single approaches.

%ADD: discussion about what is good level of classification?
%poor/ acceptable / good accuracy?

\subsection{Miss rate}

\begin{figure}
	\centering
	\includegraphics[width=1\columnwidth]{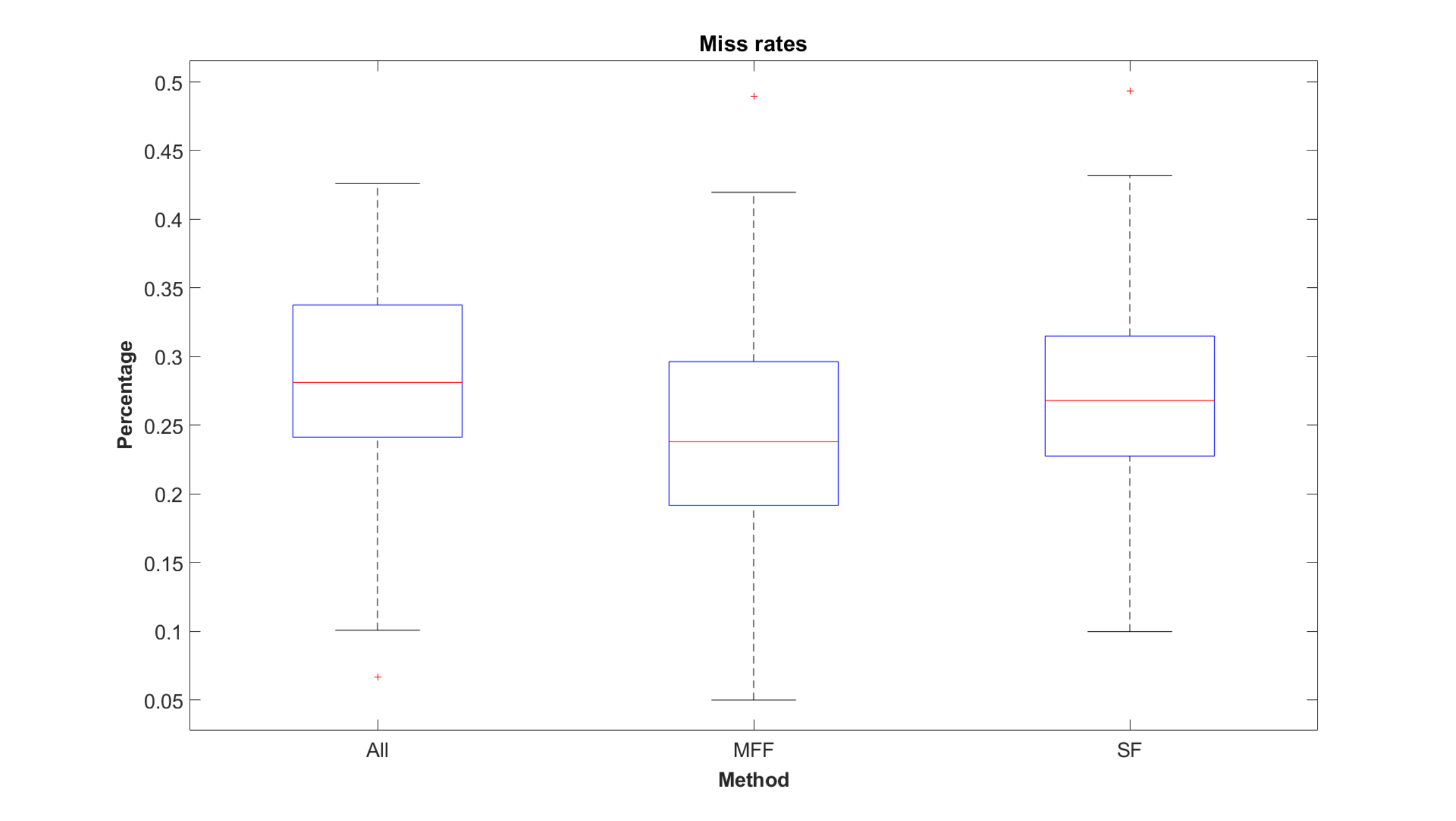}
	\caption{Miss rates of ternary methods. }~\label{fig:missRates}
\end{figure}

The miss rate is a metric that measures the rate of wrongly classified samples belonging to class x, but predicted to belong to class y. The models are better at predicting the membership of samples belonging to expert and intermediate classes than the novice class. This results in miss rates that are only little lower than chance level when looking at the median miss rates (All: 28.12\%, MFF: 23.81\% and SF: 26.80\%, see Fig ~\ref{fig:missRates}). The upper adjacent shows a high range of miss rates reaching even values of over 43.19\% for the SF-model. The MFF-model has the lowest median miss rate of all three methods with a miss rate of 41.96\%. 

\subsection{Recall}

\begin{figure}
	\centering
	\includegraphics[width=1\columnwidth]{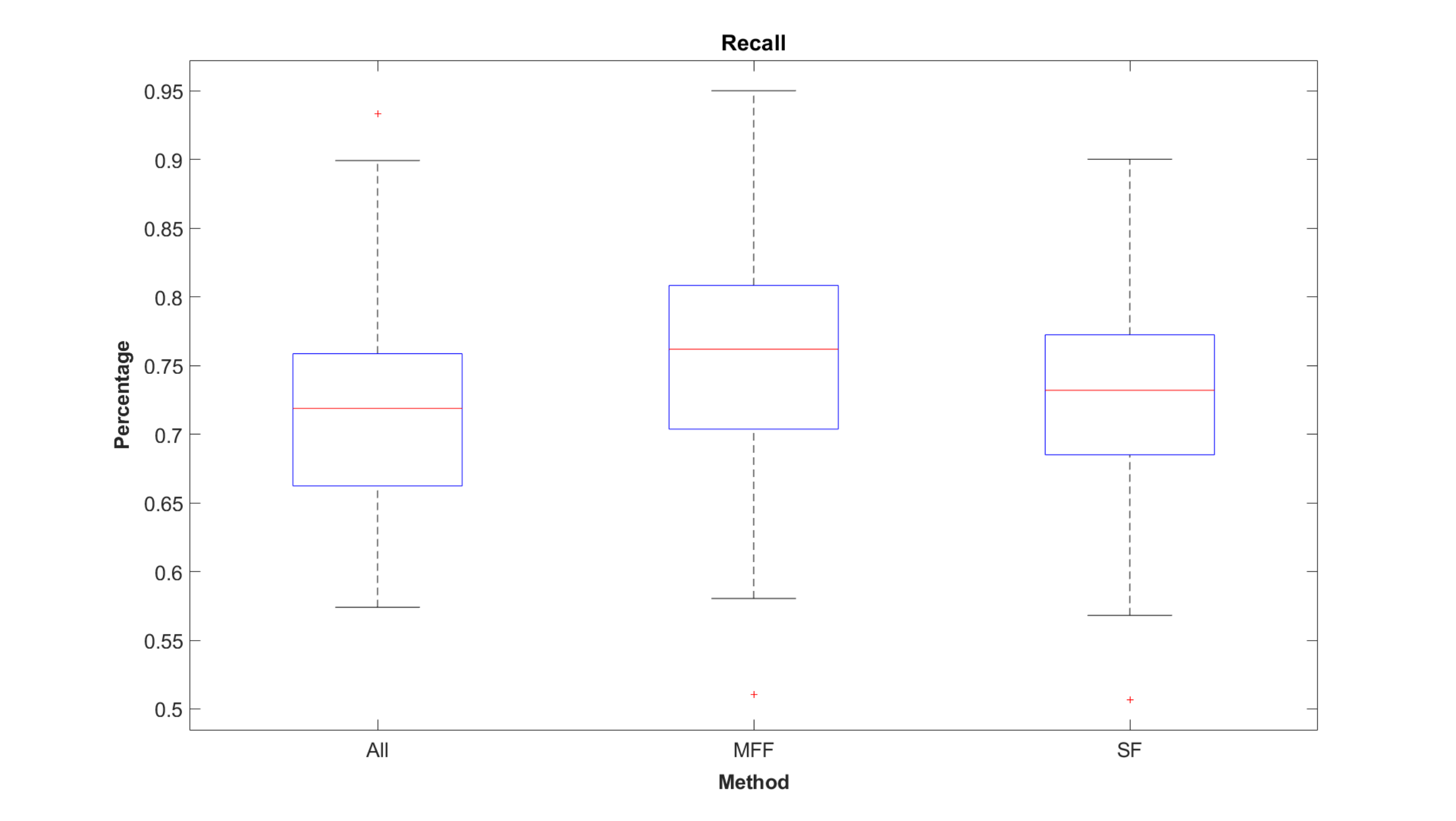}
	\caption{Recall values of ternary methods. }~\label{fig:recalls}
\end{figure}

Recall provides information about the rate of predicted samples belonging to class x in relation to the number of samples that really belong to class x. All three models have a median recall of over 70\% (as can be seen on Fig ~\ref{fig:recalls}). In the ternary case, chance level is at 33.33\% which means all models have a recall over two times higher than chance level as the lower adjacent of all three models is higher than 33.33\%. The MFF-model median is the highest at 76.18\% followed by the SF-model at 73.194\% and the ALL-model at 71.87\%. Again the MFF-model has the best performance values of all three methods.

%\subsubsection{F1-Scores}

%As a weighted average of precision and recall, the F1-score takes both false positive as well as false negatives into account. Therefore, f1-score is well suited to gain information about the overall performance of a model. The median value for f1-score range between 36.5\% (ALL-model) to 37.8\% (MFF-model). Again the MFF-model shows the best performance of all three methods. All of the f1-scores are close to chance level which 

%\begin{figure}
%	\centering%
%	\includegraphics[width=1\columnwidth]{figures/f1scores.png}
%	\caption{F1-scores for ternary methods. }~\label{fig:f1scores}
%\end{figure}

\subsection{Most frequent features}

The most frequent features in 100 runs are summarized in Table \ref{tbl:mff}. Only the minimum of the saccade duration has $p > 0.011$. This means the differences are not statistically significant. All other features show significant differences, signifying that a Mann-Whitney-U-test discards the null hypothesis that there are no differences with $p < 0.011$ for each of the features.

\begin{table}[!ht]
	
	\begin{adjustwidth}{-2.25in}{0in} % Comment out/remove adjustwidth environment if table fits in text column.
		\centering
		\caption{
			{\bf All most frequent features.}}
		\begin{tabular}{|l|c|c|c|c|c|c|}
			
			\hline
			\multicolumn{7}{|c|}{\bf Most frequent features} \\  \hline
			
			Features & derivation & novice & intermerdiate & expert &  p-value & hypothesis discarded  \\
			
			\thickhline
			saccade duration (ms) & std. dev. & 38.869 & 59.726 & 35.548 & 3.33*e-08 & 1\\ \hline		
			saccade duration (ms) & minimum & 26.514 &  26.127 & 25.019 & 0.242216408  & 0\\ \hline
			peak saccade deceleration (\textdegree$/s^2$) & std. dev.  & 4166.262 & 4770.063 & 5492.287 & 2.49*e-18 & 1\\ \hline
			peak saccade velocity (\textdegree$/s$) & std. dev.  & 129.294& 136.529 & 157.740 &  6.19*e-07 & 1\\ \hline
			smooth pursuit dispersion (pixels) & average  & 622.805 & 425.089 & 399.939 & 9.66*e-82 & 1\\ \hline			
			smooth pursuit dispersion (pixels) & minimum  & 185.437& 168.320 & 336.016 & 5.44*e-12 & 1\\ \hline			
			smooth pursuit dispersion (pixels) & maximum  & 1085.903& 694.370 & 505.031 & 1.52*e-81 & 1\\ \hline			
			
		\end{tabular}
		\label{tbl:mff}
		\begin{flushleft}\end{flushleft}
		
	\end{adjustwidth}		
\end{table}

\section{Discussion}
\label{sec:disc}
%Discussion

%1.	Restate research question:
%2.	Wie gut hat es funktioniert?
%a.	Was sind die Vorteil?
%b.	Was sind die Nachteile?
%c.	Wo sind stärken und schwächen?

%1.	Restate question
In this work, we have presented a diagnostic model to classify the eye movement features of soccer goalkeepers into expert, intermediate and novice classes. We further investigated how well the features provided by the diagnostic model led to explainable behaviour.

%2.	Wie gut war das Model?
Our model has shown that eye movement features are well suited to classifying different expertise classes. With a support vector machine, we were able to distinguish three different expertise groups with an accuracy of 78.2\% which is considered a good quality for current machine learning techniques. As the performance values differ, the real world application is questionable until we can involve more participants and provide a more robust system.

According to the miss rates the following statement can be made: the model can distinguish correctly between experts and intermediates. This is due to the fact that experts and intermediates have already been tested in the sense that they play in higher leagues and have already proven their ability. Thus, there is a ground truth for these classes.  A weakspot of the classification is the novice group. Since the novice group consists of participants with no regular training or involvement in competitions, novices can be equally talented players regarding their gaze behaviour who have simply not yet proved their ability in a competition. This thesis is especially evident in the false negative rate of 1.6\% and the false positive rate of 18.6\% from the binary model.  This means 18.6\% of novice samples are classified as intermediate samples, but only 1.6\% of the intermediate samples are classified as novice.
As is usual in expertise research, a proportion of low performers (novices) can also be found in higher classes. Our models confirm that the correct classification of novices is considerably more difficult than other classes since there is, to date,no objective ground truth.

Still, our model achieved a very good average accuracy of 78.2\%. Most likely, a model with more subjects and a finer graduation of the novices would offer a much better result. Machine learning models are data-driven and therefore can learn more from more data. However, the number of elite youth goalkeepers in Germany who can provide samples for the expert class is highly restricted. An additional step would be to define a more robust ground truth for participants classified as novices. As it is more important that the model does not downgrade participants with higher expertise to a lower class, it can still be used as a diagnostic model. As aforementioned, the false positive rate only shows, that some novices with limited experience can perform better than others and therefore be classified into a higher class. This is correct because their gaze behaviour is closer to intermediates than it is to typical novices.

%3.	Wie gut konnte man unterschiede erklären?

By examining the individual eye movement features in more detail we have shown that, on the one hand, a subset of features is sufficient to create a solid classification and, on the other hand, that the differences in eye movement behaviour between the individual groups are difficult to interpret. 
We only investigated  the most frequent features since these features built the best performing model. The differences are noticeable, but hard to interpret as there is no simple characteristic behind these features. 

%3.1	saccade std
There are indications that 1) experts (std. dev. 35.54 ms) as well as novices (std. dev. 
38.86 ms) have a more homogeneous saccade behaviour compared to intermediates (std. 
dev. 59.72 ms). The lengths of the saccades differ less. However, it would be a fallacy to 
attribute the same viewing behavior to novices and experts due to the similar 
standard deviation and minimum duration of the saccades (novice: 26 ms, intermediate: 
25 ms, expert: 25 ms). It is clear that both groups have similarly long saccades, but the novices have similarly long saccades and the experts similarly short saccades. 

%3.2 Fixation duration
Conversely, this means that the experts might have longer fixations than the 
novices and intermediates. These findings are in line with Mann et al. [20] who show 
that experts are overrepresented in fewer, but longer fixations. Their visual strategy is 
often based on longer fixations to avoid saccadic suppression (which might lead to 
information loss). In our statistics, fixation durations did not exhibit to have 
significant differences between the three groups. Which is in line with the findings of 
Klostermann et al. \cite{klostermann2020fewer}. It also might be based on the split of the fixation values in short fixations and smooth pursuits. The source of these differences may also be the age difference between the single groups (see Table 1). With the current data, this is not rigorously answerable.

%3.3 Maximum peak deceleration
Further differences between the groups can be found in the maximum peak 
deceleration of the saccades. There is a continuous increase in the maximum 
deceleration speed of the novices' saccades(4166.262 \textdegree$/s^2$) to intermediates(4770.063 
\textdegree$/s^2$) to experts (5492.287\textdegree$/s^2$), which is in line with the findings of Zwierko et al. \cite{zwierko2019oculomotor} who say that the deceleration behaviour can be inferred from different expertise classes. 

%3.4 Smooth pursuit
One observation made by the experimenter during the study was that novices often follow 
the ball with their gaze for a long time. This behavior is less evident among experts. 
They tend to only look at the ball when it has just been passed or when they themselves 
are not in play. At these times, the ball can not change its path. This observation is 
supported by the values of the smooth pursuit dispersion. With 505.031 pixels 
maximum and 336 pixels minimum, experts have a very narrow window of smooth 
pursuit lengths. Basically, the maximum smooth pursuit of the experts (505.03 pixels) is 
less than half as long as the novices (1085.90 pixels) and the minimum smooth pursuits 
(expert: 399 pixels, intermediate 425 pixels, novices 622 pixels) is still 1/3 shorter than
the novices. The intermediates are placed in the middle between the two groups. Again,
the values are continuously decreasing. Based on the continuity of the average smooth 
pursuits that correlate negatively with the classes, as well as the maximum and 
standard deviation, it can be concluded that experts tend to make smooth pursuits of a 
more regular length. One explanation for this could be that, in addition to the 
opponents and players, the ball, as an almost continuously moving object, attracts a 
high level of attention. In order to maintain a clear overview in the decision-making 
process, soccer players are taught the following behavior: Shortly before the ball arrives 
at the pass goal, you look at it. This is done until the ball is passed away. Since the 
path of the ball can only be changed by a player who is in possession of the ball and not 
in the middle of a pass, it is only necessary to follow the path of the ball at the 
beginning and end of the pass. In the meantime, players should scan the environment for changes to keep track of options in the field. This leads to short smooth pursuits 
around the ball before the end and at the beginning of each pass so that experts can 
appreciate the ball and follow the ball with similarly long smooth pursuits. On the other 
hand, as aforementioned before, novices often follow the ball's path almost continuously or, at least, very often. The characteristics of the smooth pursuit support this theory. 
The characteristics of smooth pursuits differ significantly from one another in the 
three groups with an average, minimum and maximum significant p-value of less than 
$1*10^-12$. The novices with 622.81 pixels make, on average, much longer smooth pursuits 
than the intermediates (525.09 pixels) and significantly more than the experts (399.93 pixels). 
With 185.44 pixels, the shortest smooth pursuits of the novices are smaller than those of 
the intermediates (168.32 pixels) and the experts with 336.01 pixels. The maximum 
values show a uniform behaviour. With 1085.9 pixels, the novices have the highest 
maximum values after the intermediates with 694.37 pixels and the experts with 505.03 
pixels. 

Although the standard deviation of the lengths of the smooth pursuits does not 
belong to the MF-features, clear differences can be seen here as well. The dispersion of 
the smooth pursuits with 201.27 pixels scatters far more among the novices than among 
the intermediates (124.85 pixels) and experts (112.41 pixels). These findings lead us to believe that a stimuli oriented investigation on gaze distribution for expertise 
recognition might reveal even more pronounced differences, i.e correlation between ball 
movement and smooth pursuits.

\subsection{Conclusion and Implications}

After the ternary classification of expertise, the next step should be the evaluation of a more robust classification model. As machine learning techniques are data-driven, adding more subjects to each group should, presumably,  provide better results. As soon as a robust model is built, a finer grained gradation should be considered to achieve a more sensible model that allows for the classification of participants in more classes by predicting their class in a more nuanced fashion. In our further work, we plan to expand our data set to more subjects in the current groups, add more nuanced classes and add a physical response mode to infer speed and correctness in a standardized, controllable and objective manner, thus increasing the immersion. however, a fully interactive mode will only be possible when CGI can provide high enough quality and cost-efficient environments. Another step is to focus on the research of person-specific, gaze-based expertise weakness detection. As soon as a robust model is achieved, another point is to integrate the model into an online diagnostic system. To use the model online, the gaze signal can be directly drawn online at 250 Hz from the eye tracker by using the provided API of the vendor. Using a multi-threaded system, the data preparation and feature calculation can be done directly online in parallel to data collection. Only the higher level features (e.g. std. deviations) need to be computed when the trial ends and fed as feature vector to the already trained model in order to estimate the class of the current trial. As predicting is completed by solving a function, the prediction result is supposed to be available few moments after the trial ends. This is necessary as the prediction is the input for the adaption of the training. This work will be implemented in an online system for real-time gaze-based expertise detection in virtual reality systems with an automatic input for the presentation device to ensure dynamic manipulation of a scene's difficulty. With a prototype running in VR, we are planning to expand the system to be used in-situ with augmented reality-glasses (AR). This may further pronounce the differences and lead to even better classifications. A more sensible model would allow, by mapping expertise on a larger number of classes, the dynamic manipulation of the difficulty level of an training system exercise or game level in virtual environments. Next to a training system for athletes and other professional groups, the difficulty level in a VR game can be dynamically adjusted based on the gaze behavior of the user. We are, however, aware that the small sample size restricts potential conclusions that can be drawn and may lead to contentious results. Another limitation of this work is the restriction presented by head movement unrelated eye movement features and the absence of a detailed smooth pursuit detection algorithm, which might be important. Therefore, in our future work we will implement an appropriate event calculation method i.e. based on the work of Agtzidis et al. \cite{agtzidis2019ground}.
This work, however,  strengthens the assumption that there are differences between the gaze behavior of experts, intermediates and novices, and that these differences can be obtained through the methods discussed.

%4.	Virtual realitys training systems (höheres Ziel)
Using machine learning techniques on eye tracking data captured in a photo-realistic environment on virtual reality glasses can be the first step towards a virtual reality training system (VRTS). Objective expertise identification and classification leads to adaptive and personalized designs of such systems as it allows for a definition of certain states in a training system. A VRTS that can be used at home and, based on its objective and algorithmic kind, allows for self-training at home.  The choice of difficulty can be adapted based on the expertise of the user. For higher skilled users, the level of difficulty can be raised by pointing out fewer cues or showing more crowded, faster/more dynamic scenes to increase the pressure placed on decisions. With enough data, it is also possible to adapt the training level based on personal deficiencies discovered during expertise identification in a diagnostic system. This can result in a system that knows a user's personal and perceptual weak spots to provide personalized cognitive trainings (e.g. different kinds of assistance like marking options, timing head movements, showing visual and auditive cues). Such a system is also potentially applicable in AR as the findings on the photo-realistic VR setup can be used in AR settings (i.e. in-situ). For uses such as AR-trainings - that can enhance physical trainings - the fundamental findings must be based on real gaze signals.
As a second step, training systems can be developed based on the diagnostic findings. As, in addition to physical training, perceptual-cognitive training forms are increasingly being researched \cite{appelbaum2018sports,wilkins2020early,burris2020visual,klemish2018visual}.

\nolinenumbers

% only for the first time
\bibliography{thebibliography}

\section*{Supporting information}

% Include only the SI item label in the paragraph heading. Use the \nameref{label} command to cite SI items in the text.
\paragraph*{S1 Video}
\label{S1_Vid}
{\bf Example video of experimental VR environment.} The video first shows the view at the inside of the sphere. With the head of the participant and the field of view. There is an example gaze signal jumping between players. A real stimulus is played on the inside of the sphere. Later the camera zooms out of the sphere to show how the video is projected on its inside.
\end{document}